\documentclass[twocolumn,showpacs,preprintnumbers,amsmath,amssymb]{revtex4}
\usepackage{graphicx}% Include figure files
\usepackage{dcolumn}% Align table columns on decimal point
\usepackage{bm}% bold math
\usepackage{color}

\begin{document}

%\preprint{CONCEPT $\#$10}

\title{Tuning the Gap of a Superconducting Flux Qubit}% Force line breaks with \\

\author{F.G. Paauw}
\author{A. Fedorov}
\author{C.J.P.M Harmans}
\author{J.E. Mooij}
 \email{j.e.mooij@tudelft.nl}

\affiliation{Kavli Institute of Nanoscience, Delft University of Technology, PO Box 5046, 2600 GA Delft}
\date{\today}% It is always \today, today,
             %  but any date may be explicitly specified

\begin{abstract}
We experimentally demonstrate the \textit{in situ} tunability of the minimum energy splitting (gap) of a superconducting flux qubit by means of an additional flux loop. Pulses applied via a local control line allow us to tune the gap over a range of several GHz on a nanosecond timescale. The strong flux sensitivity of the gap (up to $\sim$0.7 GHz/m$\Phi_0$) opens up the possibility to create different types of tunable couplings that are effective at the degeneracy point of the qubit. We investigate the dependence of the relaxation time and the Rabi frequency on the qubit gap.
\end{abstract}

% PACS, the Physics and Astronomy
                             % Classification Scheme.
%\keywords{Suggested keywords}%Use showkeys class option if keyword
                              %display desired
\maketitle
Superconducting circuits are promising candidates for the implementation of scalable quantum information processing~\cite{review}. To this purpose it is important to be able to selectively couple arbitrary quantum bits. Coupling multiple quantum two-level systems to a harmonic oscillator promises to be a successful strategy to create selective quantum gates in quantum optics and atomic physics~\cite{QO_ions}, and in superconducting charge~\cite{CPB_cavity} and phase qubits~\cite{phase_cavity}. In superconducting flux qubits single-qubit rotations~\cite{Delft_Rabi}, tunable qubit couplings~\cite{fqubit_tunable_coupling} and two-qubit quantum gates~\cite{2qubit_gate} have been demonstrated as well as coupling between a flux qubit and a harmonic oscillator~\cite{fqubit_HO,fqubit_HO2}. To controllably couple qubits via a harmonic oscillator bus requires the ability to tune the qubits in and out of resonance.

In this Letter, we demonstrate the implementation of an additional flux loop to vary the minimum energy splitting, called the gap, of a superconducting flux qubit. In principle this control allows a fast change of the qubit resonance frequency while remaining at the point where the coherence properties of the qubit are optimal, i.e. at the gap. The large coupling makes this tunable qubit a good candidate to implement different types of qubit coupling besides $\sigma_z\sigma_z$, such as $\sigma_x\sigma_z$ and $\sigma_x\sigma_x$~\cite{Kerman08}.

\begin{figure}
\includegraphics[width=83mm]{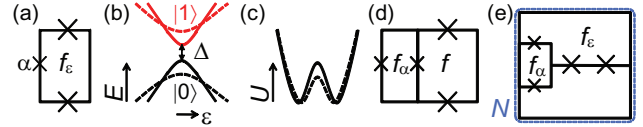}% Here is how to import EPS art
\caption{\label{fig:scheme}(color online). (a) The flux qubit with $\alpha$-junction. (b) The ground state $|0\rangle$ (black) and excited state $|1\rangle$ (red) energy levels of the flux qubit vs magnetic energy of the qubit $\epsilon$. As the $\alpha$-junction becomes smaller (dashed line) the gap $\Delta$ increases and the persistent current $I_p$ (slope) decreases. (c) The qubit double well potential. (d) Simplest implementation of a tunable-$\alpha$ qubit. The SQUID loop on the left serves as a tunable junction. (e) Tunable-$\alpha$ circuit used in this experiment. The 8-shaped gradiometric topology of the $\epsilon$-loop allows for independent control of $\alpha$. The gradiometer loop is used to trap N fluxoids to pre-bias the qubit at its degeneracy point ($\epsilon$=0).}
\end{figure}

Commonly, a flux qubit consists of a small inductance superconducting loop intersected by three Josephson junctions (Fig.~\ref{fig:scheme}(a))~\cite{Mooij99}. If the flux penetrating the loop is close to half a superconducting flux quantum $\Phi_0/2$ (mod $\Phi_0$), with $\Phi_0=h/2e$, the two lowest energy eigenstates can be used as a qubit (Fig.~\ref{fig:scheme}(b)). The qubit is characterized by the gap $\Delta$ and by the persistent current $I_p$. The energy eigenstates are linear combinations of clockwise and counterclockwise persistent-current states. In the persistent-current basis the qubit Hamiltonian can be written as
\begin{equation}
{H}=-
\frac{1}{2}%
(
 \epsilon\sigma_z+\Delta\sigma_x
) \label{hamiltonian},
\end{equation}
where $\epsilon=2I_p(f_\epsilon-\frac{1}{2})\Phi_0$ is the magnetic energy bias, with $f_\epsilon$ the magnetic frustration of the qubit loop. $\sigma_{z}$ and $\sigma_{x}$ are Pauli spin matrices.

The qubit gap is determined by the critical current of the three junctions and their capacitance. Two larger junctions have critical current $I_c$, that of the third junction is smaller by a factor $\alpha$ ($0.5<\alpha<1$). Around $f_\epsilon$=0.5, the system has a double well potential (Fig.~\ref{fig:scheme}(c)), each well connected with a persistent-current state. The barrier between the wells depends strongly on $\alpha$. When $\alpha$ is reduced, the barrier becomes smaller and the coupling $\Delta$ between the two circulating current states $\pm I_p$ increases (dashed lines Fig.~\ref{fig:scheme}(b)). At $\alpha$=0.5, the barrier vanishes. The values of the junction parameters $I_c$, $\alpha$ and the capacitances are fixed in fabrication. As $\Delta$ depends exponentially on these values, small variations, in particular of $\alpha$, lead to large variations in $\Delta$. The persistent current $I_p$ only varies slowly with $\alpha$.

\begin{figure}
\includegraphics[width=83mm]{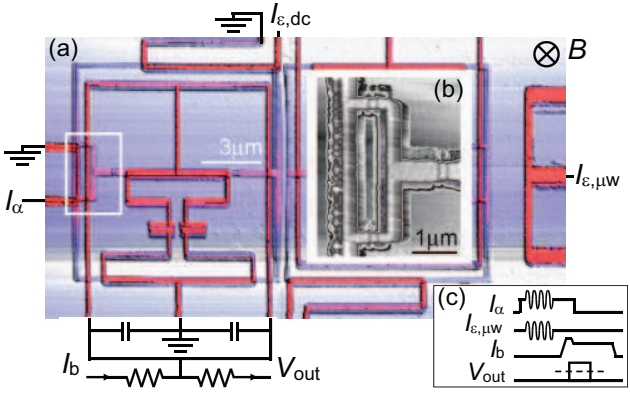}% Here is how to import EPS art
\caption{\label{fig:afm}(color online). (a) Atomic Force Micrograph (AFM) of the sample. The 8-shaped (12 $\mu m)^2$ gradiometer qubit is shown in blue with centered on the left the $\alpha$-loop (white box). The readout and control circuitry in the top layer are shown in red. Both layers are fabricated of aluminum using electron beam lithography and two-angle shadow evaporation. The inset (b) shows the $\alpha$-loop in a separate AFM graph and without the SQUID layer. Its area has been minimized to reduce crosstalk to the control circuitry. The two $\alpha$-junctions are designed to be half the size of the larger junctions. The inset (b) covers a second tunable-$\alpha$ qubit that is turned off ($N=0$) in this experiment. (c) Schematic representation of control and measurement pulses.}
\end{figure}
A tunable gap can be implemented by replacing the $\alpha$-junction by two parallel junctions (with critical current $I_{c\alpha}$) forming a SQUID (left loop Fig.~\ref{fig:scheme}(d))~\cite{Mooij99}. Manipulation of the magnetic frustration $f_\alpha$ in this $\alpha$-loop changes the critical current of the SQUID and therefore the effective $\alpha$ value, following $\alpha=2I_{c\alpha}|\cos(\pi f_\alpha)|$. This technique is common in charge qubits~\cite{Vion02}. It has also been used in flux qubits with large inductance and a single Josephson junction~\cite{Friedman00}, and in a large inductance phase qubit~\cite{Poletto08}. It is clear from the geometry in figure~\ref{fig:scheme}(d) that the frustration of the $\alpha$-loop $f_\alpha $ adds to the magnetic energy frustration $f_\epsilon$=$f+ \frac{1}{2} f_\alpha$ of the qubit. The range over which one needs to vary these frustrations is very different, with $\Delta f_\alpha \gg \Delta f_\epsilon$. Therefore, the crosstalk from $f_\alpha/2$ imposes strong compensation requirements on the implemented design~\cite{Shimazu08}.

To solve this problem and ensure independent tunability of $f_\epsilon$ and $f_\alpha$, we have replaced the single $\epsilon$-loop with a gradiometric, double loop (Fig.~\ref{fig:scheme}(e)). The specific symmetry of this design makes the magnetic energy frustration $f_\epsilon$ insensitive to any applied homogenous magnetic frustration and to the magnetic bias of the $\alpha$ loop $f_\alpha$. This enables one to selectively control $\alpha$, and therefore the gap $\Delta$, without changing the qubit magnetic energy frustration $f_\epsilon$ and vice versa.

The atomic force microscope graph of figure~\ref{fig:afm}(a) shows the qubit layout together with the readout and control circuitry used in our experiment. On the left the qubit is portrayed in blue (color online). The $\alpha$-loop, marked by a white rectangle, is shown enlarged in figure~\ref{fig:afm}(b). An inductively coupled switching SQUID is placed asymmetrically on top of the lower $\epsilon$-loop to read out the qubit state. It is electrically isolated from the qubit layer. This top layer also contains three current-carrying excitation lines have been deposited to allow local AC and DC control of $\alpha$ and $\epsilon$. Together with the homogeneous magnetic field $B$, generated by an external superconducting coil, they are used to control the frustrations $f_\alpha$ and $f_\epsilon$ according to\\
\definecolor{grey}{rgb}{0.5,0.5,0.5}
\begin{equation}
\left[
\begin{array}{c}
f_\alpha\\
f_\epsilon
\end{array}
\right]\
\Phi_0=
\left[
\begin{array}{cccc}
  \bf{M_{\alpha,\alpha}} & \textcolor{grey}{M_{\alpha,\epsilon_{dc}}} & \textcolor{grey}{\delta M_{\alpha,\epsilon_{\mu w}}} & \bf{A_\alpha} \\
  \textcolor{grey}{\delta M_{\epsilon , \alpha}} & \bf{M_{\epsilon,\epsilon_{dc}}} & \bf{M_{\epsilon,\epsilon_{\mu w}}} & \textcolor{grey}{\delta A_\epsilon}
\end{array}
\right]\
\left[
\begin{array}{c}
 I_\alpha \\ I_{\epsilon_{dc}} \\ I_{\epsilon_{\mu w}} \\ B
 \end{array}
\right].\label{fluxes}
 \end{equation}

Here $M_{i,j}$ is the mutual inductance between the current line $j$ and the $\alpha$- or $\epsilon$-loop ($i=\alpha , \epsilon$)~\cite{Mvalues} and $A_\alpha$ is the area of the $\alpha$-loop. The desired coupling parameters are represented in bold; the other elements are reduced as much as possible within the design.

The flux in the $\alpha$-loop and thus the qubit gap $\Delta$ can be controlled slowly by the external magnetic field $B$ and fast by the current $I_\alpha$. The latter is used to send both dc current pulses and microwave radiation pulses (Fig.~\ref{fig:afm}(c)).
The qubit frustration $f_\epsilon$ can be locally controlled by two lines. Its DC bias point is set by a current $I_{\epsilon,dc}$ through the top coil and microwaves $I_{\epsilon,\mu w}$ are coupled in via the antenna on the right.
Both lines on the left and right are connected to 50 $\Omega$ microwave sources. The symmetry of their design is such that $\delta M_{\epsilon,\alpha}$, $\delta M_{\alpha,\epsilon_{\mu w}}$ and $\delta A_\epsilon$ are zero to first order. This implies that $I_{\alpha}$ ($I_{\epsilon_{\mu w}}$) should be decoupled from $f_\epsilon$ ($f_\alpha$).

The solid superconducting loop of the gradiometer is used to trap fluxoids~\cite{Majer02}. With $N$ fluxoids trapped the qubit is pre-biased at $f_\epsilon = \frac{1}{2} N$. Note that N has to be odd to pre-bias the qubit at its degeneracy point ($\epsilon$=0) where it is to first order insensitive to flux noise.

Only taking into account the bold parameters of Eq.(\ref{fluxes}), the qubit Hamiltonian of Eq.(\ref{hamiltonian}) becomes
\begin{equation}
{H}=-
\frac{1}{2}%
\left\{
2I_p(B,I_\alpha)f_\epsilon(I_{\epsilon,dc},I_{\epsilon,\mu w})\Phi_0\sigma_z+\Delta(B,I_\alpha)\sigma_x
\right\}. \label{eq:ham}
\end{equation}

Qubit state transitions can be induced by applying microwave signals to the $I_{\alpha_{\mu w}}$ or $I_{\epsilon_{\mu w}}$ line, with a frequency resonant with the qubit level separation $F=(E_1-E_0)/h$ (Fig.~\ref{fig:scheme}(b)). Note from the Hamiltonian that microwave pulses sent through these two lines rotate the qubit along different axes. For resonant pulses longer than the decoherence time, the qubit transition saturates and the populations of the ground and excited state approach 50$\%$.

\begin{figure}
\includegraphics[width=83mm]{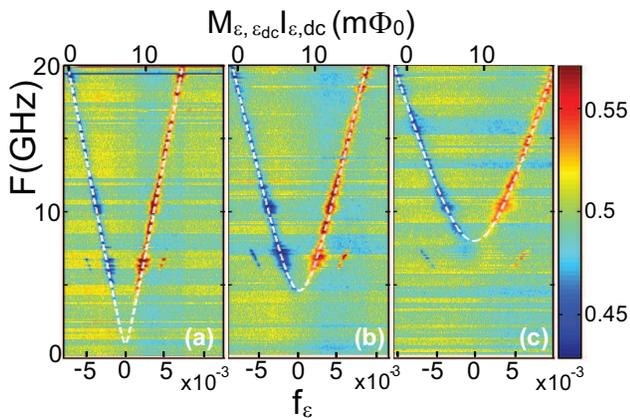}% Here is how to import EPS art
\caption{\label{fig:spectra} (color online). Microwave frequency $F$ vs magnetic frustration $f_\epsilon$ at $\alpha$ values 0.67 (a), 0.64 (b) and 0.61 (c). This change in $\alpha$ is obtained by applying a negative, zero or positive current pulse $I_\alpha$ at fixed magnetic field B. The color indicates the switching probability of the SQUID. The energy gaps and persistent currents obtain from a fit to Eq.(\ref{hamiltonian}) are ($\Delta$, $I_p$) = (1 GHz, 525 nA), (4.6 GHz, 441 nA) and (8 GHz, 345 nA) for (a), (b) and (c) respectively. In the experiment $f_\epsilon$ is varied by changing the current $I_{\epsilon,dc}$ (top axes).}
\end{figure}
Figure~\ref{fig:spectra} shows three energy spectra obtained by sweeping the qubit magnetic frustration $f_\epsilon$ at constant microwave frequency $F$. After each microwave pulse a short bias current pulse $I_b$ is applied to the SQUID. The output voltage $V_{out}$ is registered to determine whether the SQUID has switched to a finite voltage state (Fig.~\ref{fig:afm}(c)). On resonance, the mean value of the persistent current changes due to the population of the excited state. This is reflected in the altered switching probability of the SQUID (color scale). To obtain a comparable level of saturation in the spectrum, we increased the microwave power with frequency. The measurement sequence is repeated a few thousand times to improve statistics.

In figure ~\ref{fig:spectra}, the value of $\alpha$ is varied from 0.67 to 0.61 by applying a negative (a), zero (b), or positive (c), current pulse $I_\alpha$ during excitation. This data clearly demonstrates that we can achieve a qubit gap modulation from 1 to 8 GHz. This large range is consistent with our model and design. Contrary to the slow tunability of the magnetic field B, the local current line $I_\alpha$ provides fast control of the gap. Depending on the pre-bias of $f_\alpha$ by the external magnetic field $B$, we can shift the fast-tunable range to include gap values closer to zero or up to 13 GHz.

To measure the response time of the gap, we shifted the $\alpha$ pulse gradually into the microwave pulse yielding two well separated resonance peaks. The observed response time of the system was limited by the few ns rise time of the $\alpha$ pulse, but is expected to be significantly shorter.

Decreasing $\alpha$ does not only increase the qubit gap, but it also decreases the persistent current. In figure~\ref{fig:spectra}, this can be inferred from the change in slope of the spectra. Note from the top axes that there is a horizontal shift of the qubit degeneracy point for different $\alpha$ values, due to the fact that the qubit is not a perfect gradiometer ($\delta A_\epsilon \neq 0$). We attribute this mainly to a difference of approximately 10$\%$ in the critical current of the two $\alpha$-loop junctions.

\begin{figure}
\includegraphics[width=83mm]{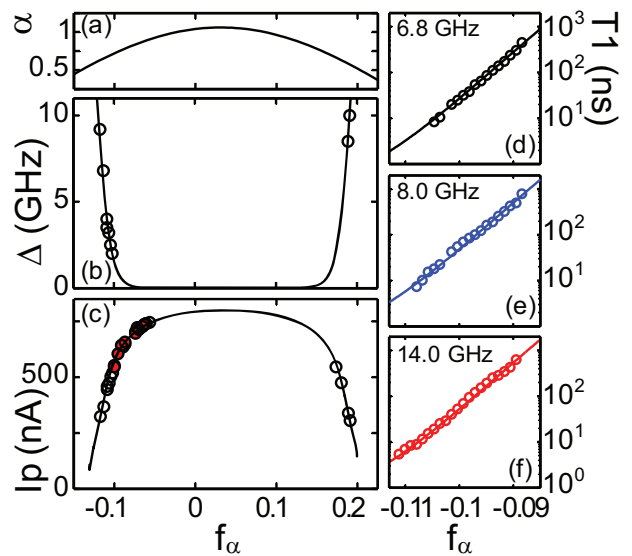}% Here is how to import EPS art
\caption{\label{fig:properties} (color online). Qubit properties $\alpha$ (a), qubit gap $\Delta$ (b), persistent current $I_p$ (c) and relaxation time $T_1$ at resonance frequency 6.8 GHz (d), 8.0 GHz (e) and 14 GHz (f) as a function of frustration $f_\alpha$. The solid lines trough the data points (circles) are obtained using the calculated parameters $C$=3 fF and $\beta$=0.063 and by fitting $I_c$=964 nA, $I_{c\alpha}$=0.53$I_c$; here $N$=1. For further details see text.}
\end{figure}

The main qubit properties as a function of $f_\alpha$ are shown in Fig.~\ref{fig:properties}. Here $f_{\alpha}$ is tuned by the external magnetic field $B$. Figure~\ref{fig:properties}(a) shows the resulting $\alpha$. Note that the trapped flux ($N$=1) imposes a phase in the $\alpha$-loop that shifts the symmetry point of $\alpha$. The individual data points for the gap $\Delta$ Fig.~\ref{fig:properties}(b) and the persistent current $I_p$ Fig.~\ref{fig:properties}(c) are obtained by fitting individual spectra to Eq.(\ref{hamiltonian}). Open and closed circles indicate whether the qubit transition was excited via $I_{\alpha}$ or $I_{\epsilon}$~\cite{dataset}.

The fit of the gap and persistent current in Figs.~\ref{fig:properties}(b,c) is obtained using the full qubit Hamiltonian beyond the two-level approximation~\cite{Orlando99}. The parameters mentioned in the figure caption are obtained in the following way: first, the critical current of the largest junction $I_c$ is fitted to the maximum persistent current. The junction capacitance $C$ is calculated from the estimated area of a Scanning Electron Microscope graph~\cite{jjcap}. The phase induced in the $\alpha$-loop by the trapped flux $N$ is inferred from the ratio of the length shared by the $\alpha$-loop over the full traploop circumference, $\beta$=0.063. Finally, the critical current value of the two $\alpha$-loop junctions $I_{c\alpha}$ is adjusted to optimize the fit.

In this experiment, the sensitivity of the gap on the $\alpha$-loop frustration $f_\alpha$ reaches a maximum of 0.7 GHz/m$\Phi_0$. This value is comparable to the $f_\epsilon$-sensitivity of a qubit with a persistent current of 100 nA. These numbers indicate that with this qubit it is possible to create both $\sigma_z$- and $\sigma_x$-type couplings and excitations. Note that the $\sigma_x$ coupling can be used at the degeneracy point ($\epsilon$=0).

The change of the gap effects the transition rate between the two qubit levels and consequently the relaxation time $T_1$. The latter was measured by monitoring the decay of the saturated resonance peak amplitude as a function of the delay time between excitation and detection. Figures ~\ref{fig:properties}(d-f) show the results as a function of $f_\alpha$ for qubit frequencies 6.8, 8 and 14 GHz. Over this range of $f_\alpha$ the gap changes from 5 GHz to 0.2 GHz.

Clearly, the relaxation time depends exponentially on $f_\alpha$ and only weakly on the qubit frequency. To understand this behavior, we simulated the effect of flux noise, critical current noise and circuit noise on $T_1$. All contributions show a similar $T_1(\Delta)$ dependence. From the spectral densities obtained from the fit we conclude that circuit noise likely is dominant. More specifically, high frequency Johnson-Nyquist noise from the on-chip resistor in the $I_{\epsilon,dc}$-line is coupled to the qubit via parasitic capacitances to the bias-line $I_{\epsilon,dc}$ and the SQUID lines; this is similar to Ref.~\cite{Steffen08}. The fit in Figs.~\ref{fig:properties}(d-f) shows the result for this noise source. The large overlap of the bias lines and the SQUID with the qubit accounts for the dominant contribution of capacitively coupled noise in this specific design. It should be stressed that much longer relaxation times have been previously obtained in flux qubits~\cite{Delft_Rabi, fqubit_tunable_coupling, 2qubit_gate, fqubit_HO, longT1}.

\begin{figure}[t]
\includegraphics[width=83mm]{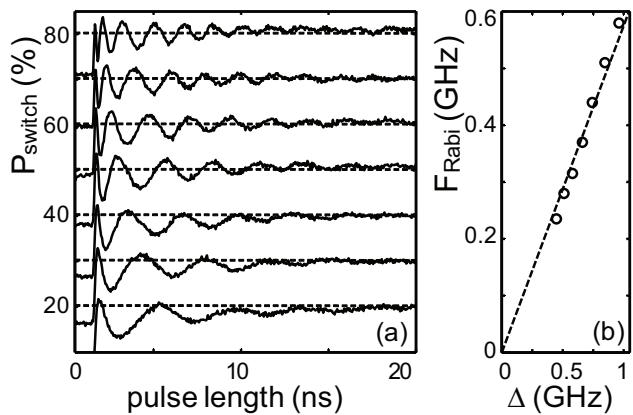}% Here is how to import EPS art
\caption{\label{fig:coherence} (a) Rabi oscillations at F=6.8 GHz and fixed power for different values of the gap ($\Delta$ is 0.45 GHz (bottom trace) to 0.95 GHz (top trace)). For clarity individual traces are shifted vertically. (b) The Rabi frequency of each trace in (a) is plotted as a function of the gap $\Delta$. The dashed line is a linear fit through the origin.}
\end{figure}

The rather strong relaxation at large gaps limits the parameter range for coherent control of this specific qubit. Figure ~\ref{fig:coherence}(a) shows Rabi oscillations at different values of the gap. The oscillations are induced at a qubit frequency of 6.8 GHz and with fixed driving power on $I_{\alpha}$. From Eq.(\ref{hamiltonian}) a linear increase of the Rabi frequency $F_{Rabi}$ versus $\Delta$ is predicted. This agrees with the experimental data (Fig.~\ref{fig:coherence}(b)). Similar Rabi dynamics is observed when driving via $I_\epsilon$. Unfortunately, measurements at the degeneracy point ($\epsilon$=0) were hampered, at larger gaps by the short relaxation time, and at smaller gaps by the temperature. Therefore, this specific sample did not allow us to demonstrate the full potential of a flux qubit with a tunable gap. The necessary improvements to increase the coherence times are within reach and are not specifically related to the implementation of the $\alpha$-loop.

In summary, we have experimentally demonstrated the rapid tunability of the energy gap of a superconducting flux qubit. This was achieved by replacing the smaller Josephson junction of the qubit with a tunable SQUID. The tunability enables one to rapidly bring the flux qubit in and out of resonance with other quantum systems, including a harmonic oscillator. These operations can be performed at the degeneracy point of the qubit, where coherence properties are optimal. In this manner, the tunable flux qubit provides an attractive component for the implementation of scalable quantum computation.

This work was supported by the Dutch Organization for Fundamental Research on Matter (FOM), EU EuroSQIP and the NanoNED program.


\begin{thebibliography}{10}
\bibitem{dataset}
The data set for $\Delta$ is smaller than for $I_p$, because below 2 GHz the gap becomes difficult to observe.
\bibitem{jjcap}
We used $C/A$=80 fF/$\mu m^2$. The capacitance of the rightmost large junction (Fig.~\ref{fig:afm}(a)) is enhanced by a capacitive shunt by the SQUID circuit, $C_{rJ}$=5.2 $C$.
\end{thebibliography}
\end{document}